%% file: main.tex
\documentclass[aps, physrev, reprint, superscriptaddress]{revtex4-2}

\usepackage[utf8]{inputenc}
\usepackage{amsmath, amssymb}
\usepackage{graphicx}
\usepackage{xcolor}
\usepackage{appendix}
\usepackage{comment}
\usepackage{braket}
\usepackage{float}

\usepackage[colorlinks=true, allcolors=blue]{hyperref}
\usepackage[capitalise]{cleveref}
\usepackage{orcidlink}

\usepackage{tikz}
\usepackage{tkz-euclide}

\begin{document}

\title{Universal flux-based control of a $\pi$-SQUID}

\author{J.~Wilson~Staples} 
\affiliation{Duke Quantum Center, Duke University, Durham, NC 27701, USA}
\affiliation{Department of Physics, Duke University, Durham, NC 27708, USA}
\affiliation{Department of Electrical and Computer Engineering, Duke University, Durham, NC 27708, USA}

\author{Thomas~B.~Smith} 
\affiliation{Centre for Engineered Quantum Systems, School of Physics, University of Sydney, Sydney, NSW 2006, Australia.}

\author{Andrew~C.~Doherty} 
\affiliation{Centre for Engineered Quantum Systems, School of Physics, University of Sydney, Sydney, NSW 2006, Australia.}

\date{\today}

\begin{abstract}
    We describe a protocol for the universal control of non-ideal $\pi$-periodic superconducting qubits.
    Our proposal relies on a $\pi$-SQUID: a superconducting loop formed by two $\pi$-periodic circuit elements, with an external magnetic flux threading the circuit. 
    The system exhibits an extensive sweet spot around half-flux where residual $2\pi$-periodic Cooper pair tunneling is highly suppressed.
    We demonstrate that universal single-qubit operations can be realised by tuning the flux adiabatically and diabatically within this broad sweet spot.
    We also assess how residual $2\pi$-periodicity in $\pi$-SQUIDs impacts holonomic phase gates.
\end{abstract}

\maketitle

\section{Introduction}\label{sec:introduction}

The last two decades has seen remarkable improvement in the coherence times and error rates of superconducting circuits~\cite{Kjaergaard_2020,Blais_2021}.
As a result, superconducting quantum computers are on the cusp of demonstrating quantum error correction below threshold~\cite{Arute_2019,Acharya_2023}.
Further progress towards fault-tolerance demands even lower error rates --- as far below threshold as possible.
Decreasing error rates below threshold substantially reduces the hardware overheard for superconducting quantum computers~\cite{Gambetta_2017}.

A new generation of protected superconducting circuits are being investigated out of a desire for qubits with much lower error rates~\cite{Gyenis_2021b}.
These circuits are engineered to have intrinsic resistance to both dephasing and depolarisation.
This is in contrast to conventional superconducting circuits, such as the transmon, which only possess partial protection against noise~\cite{Koch_2007}.

The cost of this enhanced protection is twofold.
First, protected qubits tend to rely on complex multi-node circuits, often in rather extreme circuit parameter regimes~\cite{Dempster_2014}.
Second, they require non-standard approaches to preparation, control, and measurement \cite{leroux2023catqubitinspired} --- a direct consequence of their increased protection~\cite{Brooks_2013}. 
We focus on the latter problem in this paper.

\begin{figure}
    \centering
    \begin{tikzpicture}
        \include{figures/qubit}
    \end{tikzpicture}
    \caption{
        Circuit diagram and effective tunneling amplitudes of a $\pi$-SQUID.
        (a) The circuit diagram for a $\pi$-SQUID, with external flux $\Phi$ and capacitive coupling $C_g$ to an external voltage $V_g$.
        The $\pi$-SQUID contains two non-ideal $\pi$-periodic junctions $a$ and $b$, shunted by a capacitance $C$.
        (b) Model for a non-ideal $\pi$-periodic junction, with parasitic single Cooper pair tunneling.
        (c, d) Effective tunneling amplitudes $E_{J_1}(\Phi)$ and $E_{J_2}(\Phi)$ of the $\pi$-SQUID as a function of external flux $\Phi$, for $d_1 = d_2 = 0$.
    }
    \label{fig:circuit}
\end{figure}
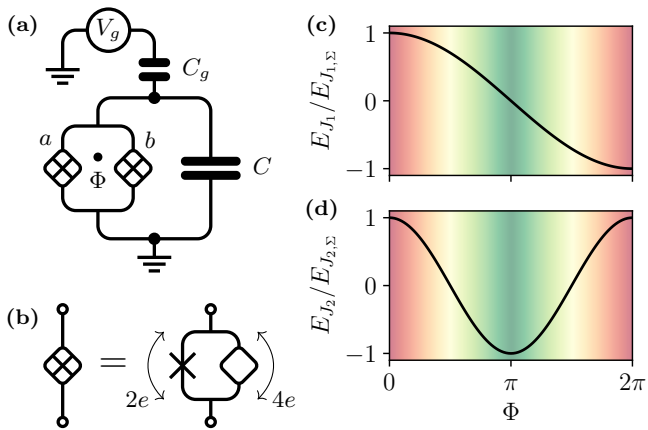

We describe a universal control scheme for protected qubits, using only the magnetic flux threaded through a pair of $\pi$-periodic circuits.
The $\pi$-SQUID setup in~\cref{fig:circuit}(a), which was introduced in Ref.~\cite{Klots_2021}, is compatible with any protected qubit that relies on an effective $\pi$-periodic junction. 
Examples include the current mirror qubit~\cite{kitaev2006protected}, the $0$-$\pi$ qubit~\cite{Brooks_2013}, the $\cos2\theta$ qubit~\cite{Larsen_2020}, and the Josephson rhombus chain qubit~\cite{Bell_2014}.

A $\pi$-junction allows for the tunneling of pairs of Cooper pairs, while completely suppressing the tunneling of individual Cooper pairs:
\begin{equation}\label{eq:pi-junction}
    \hat H = - E_{J_2} \cos{2 \hat \phi},
\end{equation}
where $\hat\phi$ is the superconducting phase across the junction, and $E_{J_2}$ is the tunneling amplitude.

A $\pi$-junction is the key circuit element for broad class of transmon-like protected qubits, where it replaces the $2\pi$-periodic potential of a Josephson junction~\cite{Smith_2020}.
Combined with a large shunting capacitance, the double-well potential of~\cref{eq:pi-junction} admits a doubly degenerate ground space, spanned by the ground states of the two wells~\cite{Groszkowski_2018}.
A qubit encoded in these degenerate states would be robust to both dephasing and depolarisation~\cite{Weiss_2019}. However, this protection also poses a problem; standard control techniques no longer work due to low-frequency qubit transitions and vanishingly small matrix elements~\cite{Maiani_2022}.
For example, the `soft' experimental realisation of the $0$-$\pi$ qubit required Raman-type control for logical operations~\cite{Gyenis_2021a}.
Though the circuit was not fully protected, conventional Rabi-style control was not feasible. 

Near-term experimental implementations of this class of protected qubits are expected have imperfect $\pi$-periodicity.
A minimal model for such a non-ideal $\pi$-junction is~\cite{kitaev2006protected,Paolo_2019,Ciaccia_2024}
\begin{equation}\label{eq:soft-pi-junction}
    \hat H = - E_{J_1} \cos{\hat \phi} - E_{J_2} \cos{2 \hat \phi},
\end{equation}
where $E_{J_1}$ is the `parasitic' Josephson energy that corresponds to the tunneling of single Cooper pairs.

In this work we describe an approach to universal control of qubits based on a $\pi$-SQUID in the presence of such parasitic single Cooper pair tunnelling. 
The flux threaded through a $\pi$-SQUID qubit leads to a controllable bias in its double-well potential~\cite{Schrade_2022,Valentini_2024}.
We show that at half-flux, the qubit has a very broad low-frequency sweet spot where $2\pi$-periodic tunneling is suppressed.
In this low-frequency regime, adiabatic and diabatic tuning of the flux allows for universal single-qubit control. 
Our discussion is at a level that is applicable to the full range of potential $\pi$-SQUID systems. 
We leave detailed noise simulations that would apply to a particular implementation to future work.

Alternative control schemes for low-frequency, protected qubits that forego resonant Rabi-driving have been studied extensively in the literature, particularly in context of spin qubits.
Spin qubits, specifically those encoded in multiple quantum dots, are controlled using baseband manipulation of their confining potential~\cite{Burkard_2023}.
A singlet-triplet spin qubit, which is encoded in a double-quantum-dot, has a voltage-tunable qubit splitting that is pulsed both diabatically and adiabatically in order to build a universal single-qubit gateset~\cite{petta2005coherent}.
In this paper, we adopt a similar approach for the control of $\pi$-SQUIDs, which also have a double-well potential, and a flux-tunable qubit splitting.
An alternative adiabatic gate scheme for $\pi$-SQUIDs based on holonomic phases was explored in Ref.~\cite{Klots_2021}, and we assess its applicability to non-ideal $\pi$-junctions.
We also note that fast, high-fidelity flux-based gates have been demonstrated for low-frequency qubits in fluxonium~\cite{Zhang_2021} and transmon-based qubits~\cite{Campbell_2020}.
In these experiments, universal control was achieved via non-adiabatic transitions, and coherent Landau-Zener interference.
These approaches are likely compatible our setup, but we will not explore them in this paper.

The structure of this paper is as follows. 
In~\cref{sec:model}, we analyse the superconducting circuit for a $\pi$-SQUID with residual $2\pi$-periodic tunneling. 
We demonstrate that the system possesses a extended sweet spot around half-flux, and that tuning an external magnetic flux allows for control over the double well potential.
In~\cref{sec:gates}, we describe how this flux-tunability allows for universal control of the $\pi$-SQUID qubit, and outline the basic principles of state preparation and measurement.
Finally, in~\cref{sec:holonomic}, we briefly analyse the quality of holonomic gates for non-ideal $\pi$-SQUIDs.

\section{The $\pi$-SQUID}\label{sec:model}

We study the superconducting circuit in~\cref{fig:circuit}(a).
This SQUID-like circuit consists of two $\pi$-periodic junctions shunted by a capacitor~\cite{Klots_2021}.
The $\pi$-junctions permit the tunneling of pairs of Cooper pairs, but are also assumed to have some parasitic Josephson effect, as indicated in~\cref{fig:circuit}(b).
Thus, we assume the Josephson Hamiltonian for the $\pi$-SQUID loop is
\begin{equation}\label{eq:josephson-hamiltonian}
    \begin{aligned}
        \hat{H}_J &= - \sum_{i = a,b} \left( E_{J_{1,i}} \cos{\hat\phi_i} + E_{J_{2,i}} \cos{2\hat\phi_i} \right),
    \end{aligned}
\end{equation}
where $\hat\phi_a$ and $\hat\phi_b$ are the superconducting phase differences across junctions $a$ and $b$. 
The parasitic tunneling of single Cooper pairs is be characterized by the parameter $E_{J_{1,\Sigma}}=E_{J_{1,a}}+E_{J_{2,b}}$, the sum of the Josephson energies in each of the junctions. 
Likewise, the tunnelling of pairs of Cooper pairs is described by the parameter $E_{J_{2,\Sigma}}=E_{J_{2,a}}+E_{J_{2,b}}$. 

Flux quantisation implies that 
\begin{equation}\label{eq:flux-quantisation}
    \phi_a - \phi_b = \Phi \mod{2\pi},
\end{equation}
where $\Phi$ is the external magnetic flux threading the $\pi$-SQUID loop, in units of $\Phi_0 = \hbar/2e$, the reduced superconducting flux quantum.
We can rewrite~\cref{eq:josephson-hamiltonian} in terms of the phase difference $\hat\varphi = (\hat\phi_a + \hat\phi_b)/2$:
\begin{equation}\label{eq:pi-SQUID}
    \hat H_J = - E_{J_1} (\Phi) \cos{\left[\hat \varphi - \varphi_0 (\Phi) \right]} - E_{J_2} (\Phi) \cos{2 \hat \varphi} ,
\end{equation}
where $E_{J_1} (\Phi)$ is the effective $2\pi$-periodic tunneling amplitude, $E_{J_2} (\Phi)$ is the effective $\pi$-periodic tunneling amplitude, and $\varphi_0 (\Phi)$ is the phase difference between the two.
The full expressions for these parameters are provided in~\cref{app:pi-SQUID-derivation}.
Like a conventional SQUID, the effective tunneling amplitudes $E_{J_1}(\Phi)$ and $E_{J_2}(\Phi)$ --- plotted in~\cref{fig:circuit}(c,d) --- are tunable via the external magnetic flux $\Phi$ threading the loop.
Unlike a conventional SQUID, there is a relative phase shift $\varphi_0(\Phi)$ between the two potentials that cannot be removed via a gauge transformation.
This relative phase shift has a significant effect on the physics of the $\pi$-SQUID.

The Hamiltonian for the full circuit is 
\begin{equation}\label{eq:circuit}
    \hat H = 4E_C (\hat n - n_g)^2 + \hat H_J,
\end{equation}
where $\hat n$ is Cooper pair number operator conjugate to $\hat\varphi$, $E_C = e^2/2C$ is the charging energy, and $n_g$ is the offset charge due to the external voltage $V_g$. 
The capacitance $C$ should be large, such that $E_C \ll E_{J_2}$.
This provides protection against dephasing, owing to suppressed charge dispersion.

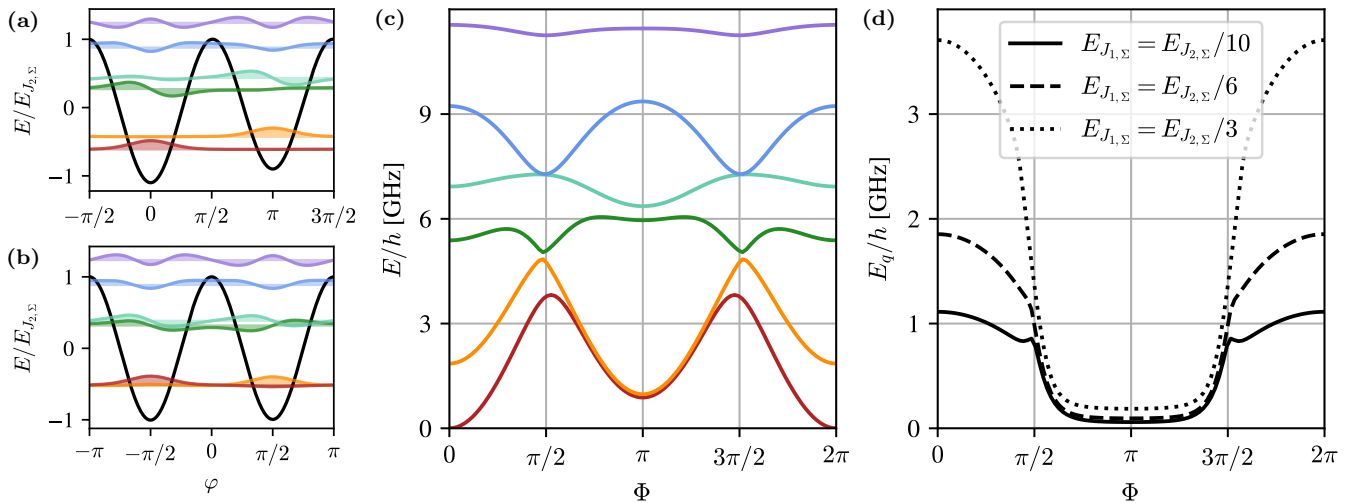
\begin{figure*}
    \centering
    \begin{tikzpicture}
        \coordinate (A) at (0-0.1, 1.95);
        \coordinate (B) at (-0.11-0.1, -1.5);
        \coordinate (C) at (5.625, 0.01);
        \coordinate (D) at (12+0.1, 0.01);
    
        \node at (A) {
            \includegraphics[
                trim={0 20 0 0}, clip,
                scale=0.825
                ]{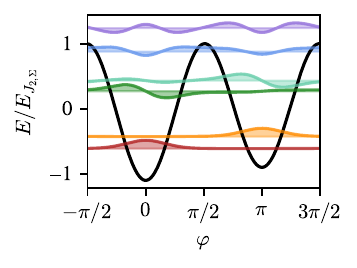}
            };
        \node at (B) {
            \includegraphics[scale=0.825]{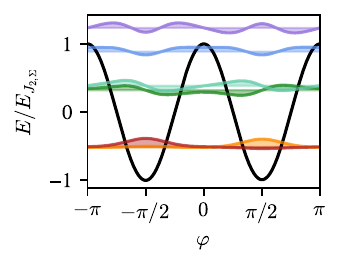}
            };
        \node at (C) {
            \includegraphics[scale=0.95]{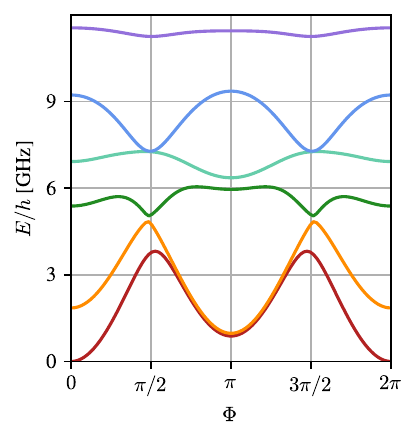}
            };
        \node at (D) {
            \includegraphics[scale=0.95]{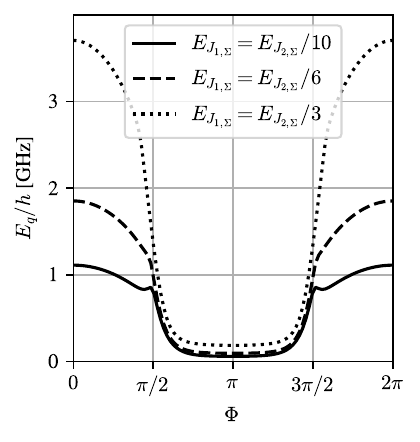}
            };

        \node at ($ (A) + (-2.15, 1+0.2) $) {\footnotesize \textbf{(a)}};
        \node at ($ (B) + (-2.15+0.11, 1.5) $) {\footnotesize \textbf{(b)}};
        \node at ($ (C) + (-2.9-0.1, 2.975+0.2) $) {\footnotesize \textbf{(c)}};
        \node at ($ (D) + (-2.825-0.15, 2.975+0.2) $) {\footnotesize \textbf{(d)}};
    \end{tikzpicture}
    \caption{
        Spectrum of a $\pi$-SQUID. 
        Calculated numerically for $E_{J_{2, \Sigma}}/E_C = 30$, $E_{J_{1, \Sigma}}/E_{J_{2,\Sigma}} = 0.1$, $E_C/h = 200$ MHz, $n_g = 0$, and $d_1=d_2=0.05$.
        (a) The first six energy eigenstates at $\Phi=0$, (b) and $\Phi=\pi$.
        (c) Energy eigenvalues as a function of flux $\Phi$. 
        (d) Splitting of the two ground states as a function of flux $\Phi$, for three different values of $E_{J_{1,\Sigma}}$.
    }
    \label{fig:spectrum}
\end{figure*}

The protection of a $\pi$-SQUID qubit depends on the relative strength of the $\pi$-periodic and $2\pi$-periodic components in the potential energy~\cref{eq:pi-SQUID}.
Ideally, single Cooper pair tunneling is fully suppressed and $E_{J_{1,\Sigma}}=0$.
In this case, the potential energy has two equal minima.
A qubit encoded in the ground states of these two wells would be highly protected against noise~\cite{Groszkowski_2018,Paolo_2019,Weiss_2019}.
Realistic devices will only have partial suppression and $E_{J_{1,\Sigma}}>0$. 
This deforms the potential, and splits the degeneracy of the ground states.
Fortunately, $E_{J_1}(\Phi)$ and $E_{J_2}(\Phi)$ have different periodicity in $\Phi$ --- see~\cref{fig:circuit}(c,d). 
Consider the region around half-flux ($\Phi=\pi$).
Here, the $2\pi$-periodic tunneling amplitude $E_{J_1}(\Phi)$ is suppressed and the $\pi$-periodic tunneling amplitude $E_{J_2}(\Phi)$ is at an extremum.
A $\pi$-SQUID qubit tuned to roughly half-flux will therefore partially regain its protection.
Note that asymmetry in the $\pi$-SQUID (i.e. $d_1,d_2>0$) leads to incomplete cancellation of the single Cooper pair tunneling, as evident in~\cref{eq:josephson-energies}.

The eigenstates of an asymmetric $\pi$-SQUID are shown in~\cref{fig:spectrum}(a,b), for both $\Phi=0$ and $\Phi=\pi$.
The flux provides a means to bias the double-well potential, and control the splitting between the two lowest energy states.
Note that $E_{J_2}(\Phi)$ is negative at $\Phi=\pi$, so the wells in~\cref{fig:spectrum}(a) are offset by $\pi/2$ relative to~\cref{fig:spectrum}(b).

Near half-flux, where the protection is maximised, the effective $2\pi$-periodicity $E_{J_1}(\Phi)$ is linearly dependent on $\Phi$, as shown in~\cref{fig:circuit}(c).
Therefore, we would expect the qubit to be linearly susceptible to flux-based dephasing, since flux noise enters directly through $E_{J_1}(\Phi)$.
However, when we calculate the spectrum numerically in~\cref{fig:spectrum}(c), we see a wide sweet spot around half-flux.
This broad sweet spot in the ground state splitting is persistent, even for larger values of $E_{J_{1,\Sigma}}$, as shown in~\cref{fig:spectrum}(d). Although in our plots we have chosen $d_1=d_2$ this in not required, and the behaviour is not qualitatively changed so long as $d_1$ and $d_2$ are of the same order of magnitude.



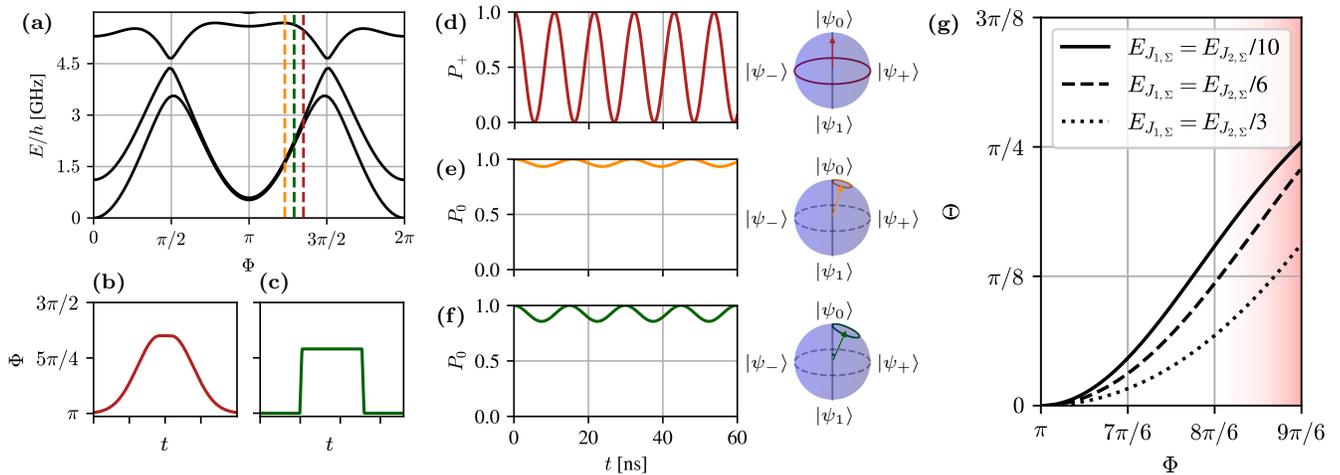
\begin{figure*}
    \centering
    \begin{tikzpicture}
        \coordinate (A) at (-6.3, 0);
        \coordinate (B) at (-0.1, 0.5);
        \coordinate (C) at (5.85, -0.02);
    
        \node at ($ (A) + (0, 1.3) $) {
            \includegraphics[scale=0.7]{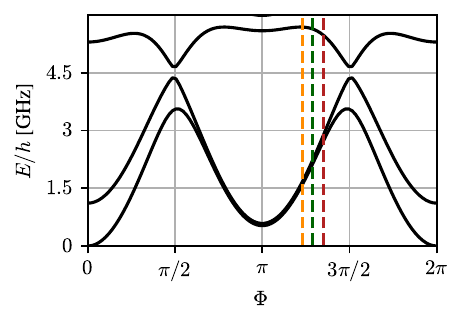}
            };
        \node at ($ (A) + (-1.31, -1.5) $) {
                \includegraphics[
                    trim={0 30 0 0}, clip,
                    scale=0.775
                    ]{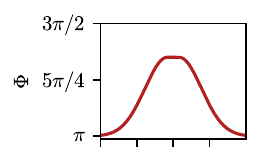}
                };
        \node at ($ (A) + (-1.31, -1.5) + (0.55, -1.2) $) {\footnotesize $t$};   
        \node at ($ (A) + (-1.31, -1.5) + (2.75, -1.2) $) {\footnotesize $t$};   
        \node at ($ (A) + (1.52, -1.5) $) {
                \includegraphics[
                    trim={45 30 0 0}, clip,
                    scale=0.775
                    ]{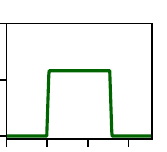}
                };
        \begin{scope}[shift=(B)]
            \node at (-1.2, 1.88) {
                \includegraphics[
                    trim={0 30 0 0}, clip,
                    scale=0.75
                    ]{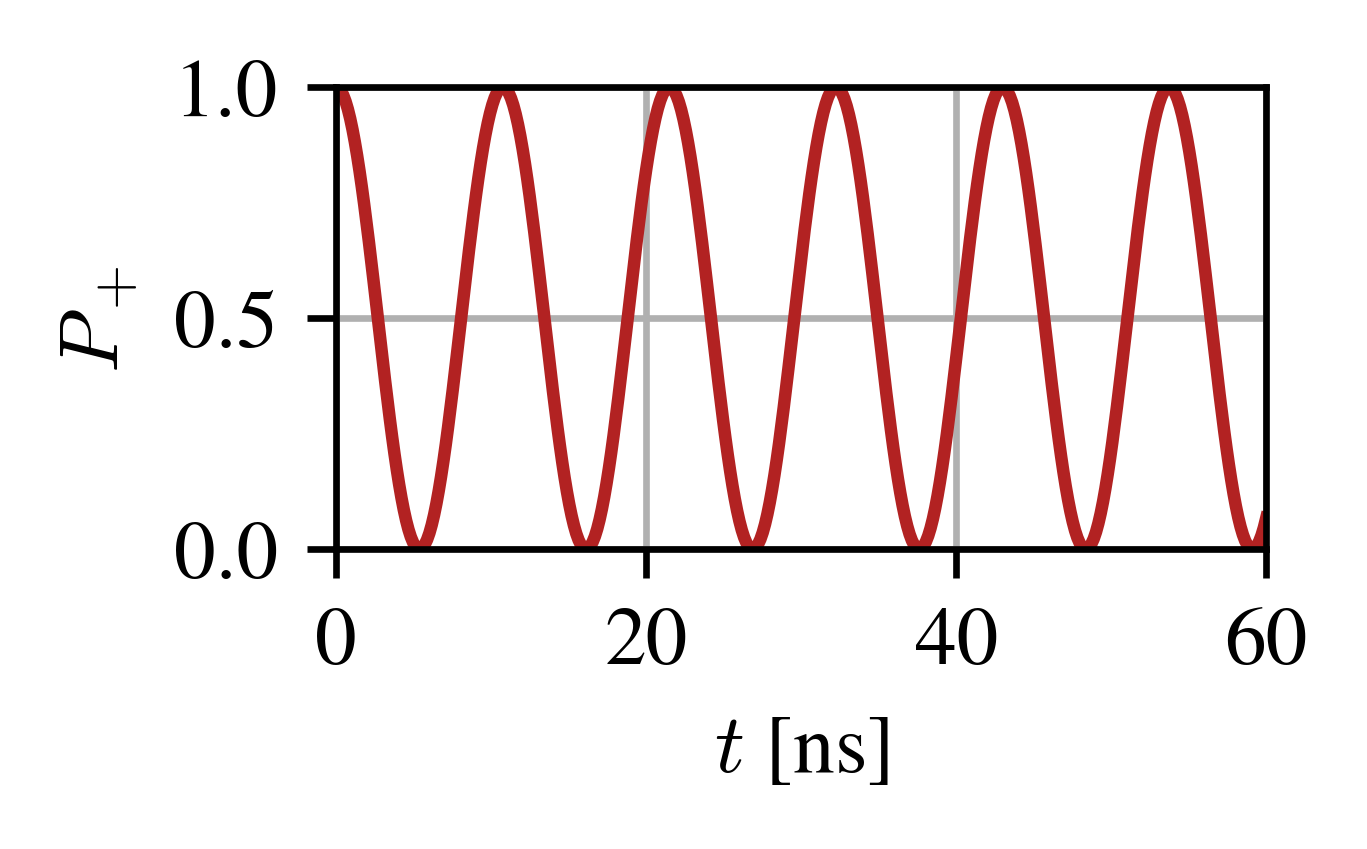}
                };
            \node at (-1.2, -0.08) {
                \includegraphics[
                    trim={0 30 0 0}, clip,
                    scale=0.75
                    ]{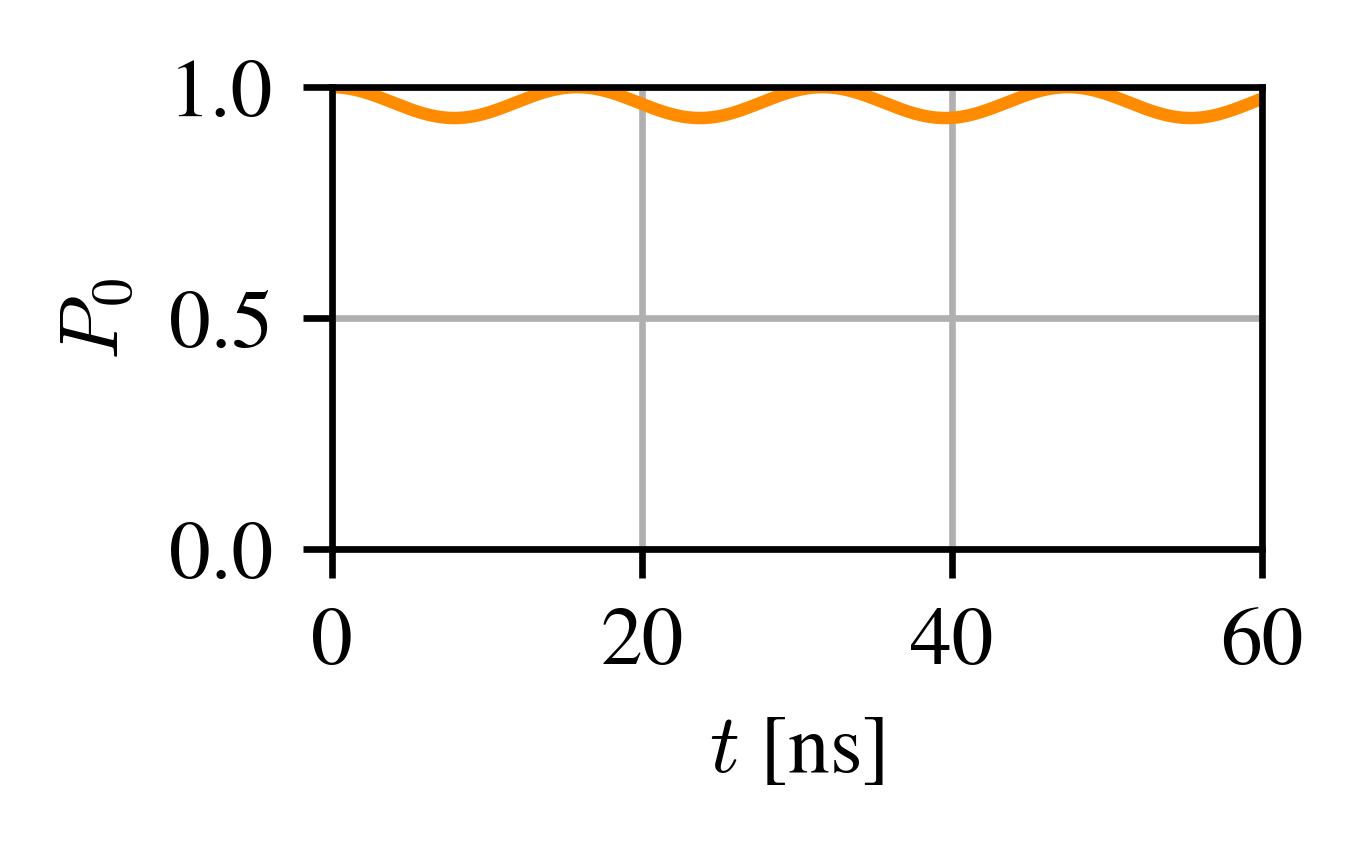}
                };
            \node at (-1.2, -2.43) {
                \includegraphics[
                    scale=0.75
                    ]{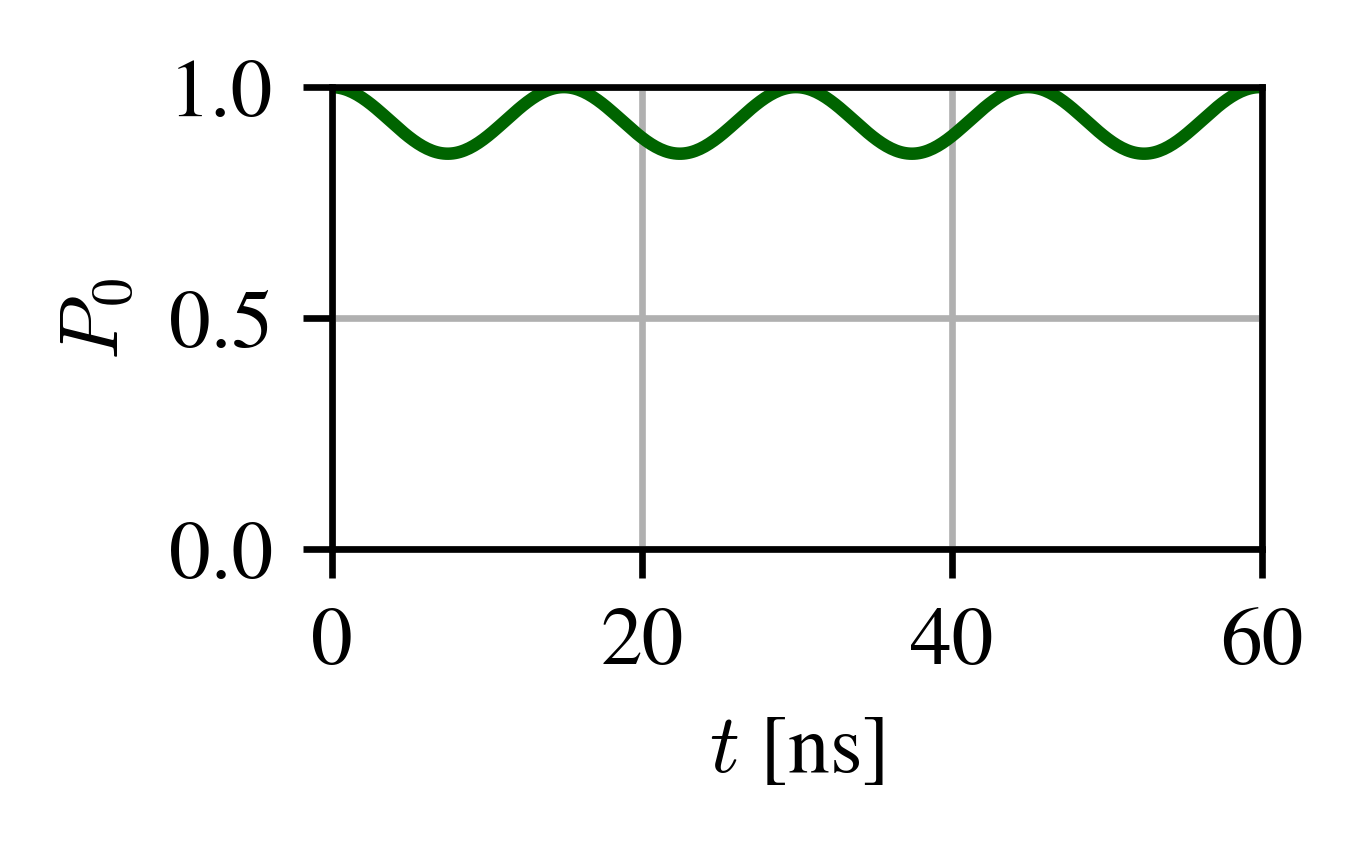}
                };
        \end{scope}
        \begin{scope}[shift=(B)]
            \node at (1.9, 1.88-0.125) {
                \includegraphics[
                    trim={29 29 29 29}, clip,
                    scale=0.45
                    ]{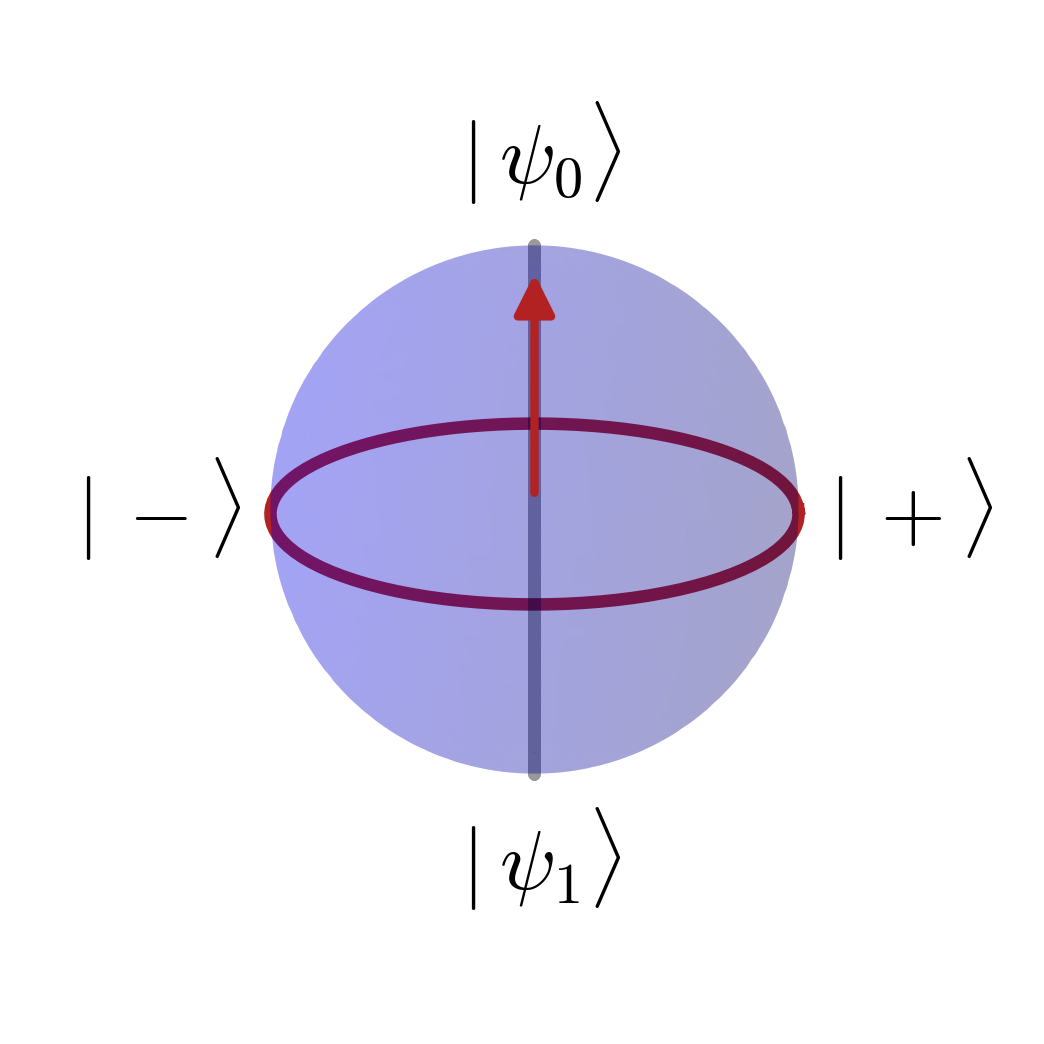}
                };
            \node at ($ (1.9, 1.88-0.125) + (0.05, 0.7) $) {\scriptsize $\ket{\psi_0}$};
            \node at ($ (1.9, 1.88-0.125) + (0.05, -0.7) $) {\scriptsize $\ket{\psi_1}$};
            \node at ($ (1.9, 1.88-0.125) + (0.9, 0) $) {\scriptsize $\ket{\psi_+}$};
            \node at ($ (1.9, 1.88-0.125) + (-0.85, 0) $) {\scriptsize $\ket{\psi_-}$};
            \node at (1.9, -0.08-0.125) {
                \includegraphics[
                    trim={29 29 29 29}, clip,
                    scale=0.45
                    ]{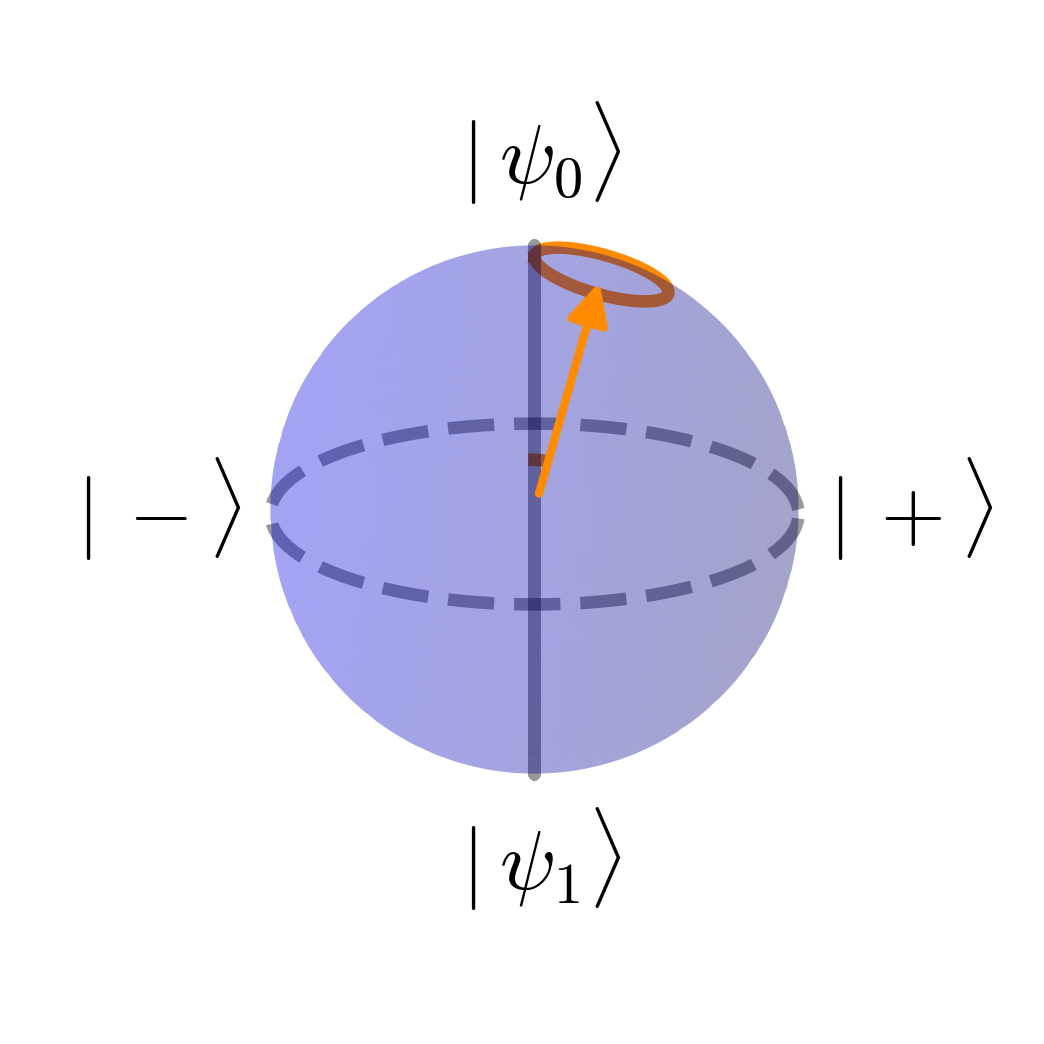}
                };
            \node at ($ (1.9, -0.08-0.125) + (0.05, 0.7) $) {\scriptsize $\ket{\psi_0}$};
            \node at ($ (1.9, -0.08-0.125) + (0.05, -0.7) $) {\scriptsize $\ket{\psi_1}$};
            \node at ($ (1.9, -0.08-0.125) + (0.9, 0) $) {\scriptsize $\ket{\psi_+}$};
            \node at ($ (1.9, -0.08-0.125) + (-0.85, 0) $) {\scriptsize $\ket{\psi_-}$};
            \node at (1.9, -2.25+0.125) {
                \includegraphics[
                    trim={29 29 29 29}, clip,
                    scale=0.45
                    ]{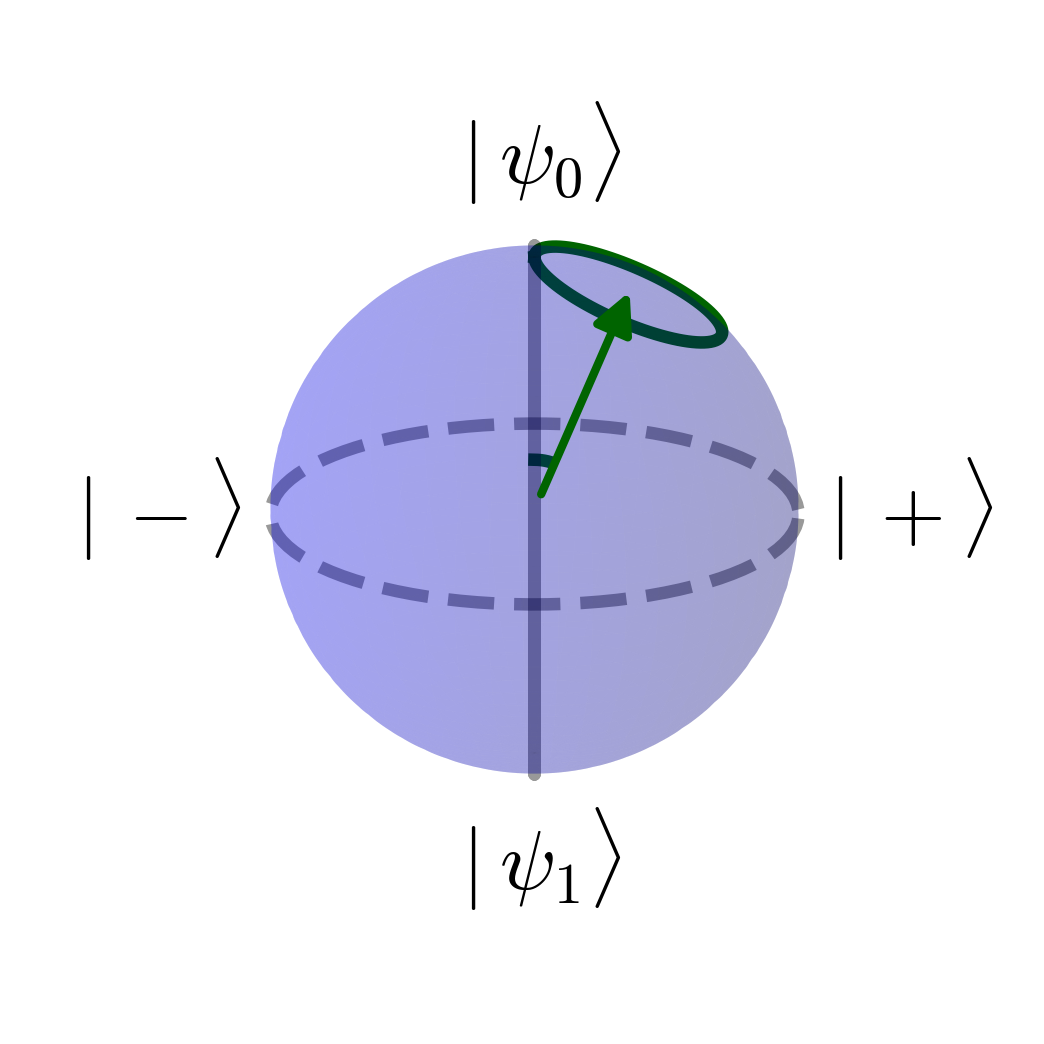}
                };
            \node at ($ (1.9, -2.25+0.125) + (0.05, 0.7) $) {\scriptsize $\ket{\psi_0}$};
            \node at ($ (1.9, -2.25+0.125) + (0.05, -0.7) $) {\scriptsize $\ket{\psi_1}$};
            \node at ($ (1.9, -2.25+0.125) + (0.9, 0) $) {\scriptsize $\ket{\psi_+}$};
            \node at ($ (1.9, -2.25+0.125) + (-0.85, 0) $) {\scriptsize $\ket{\psi_-}$};
        \end{scope}
        \node at (C) {
            \includegraphics[scale=0.88]{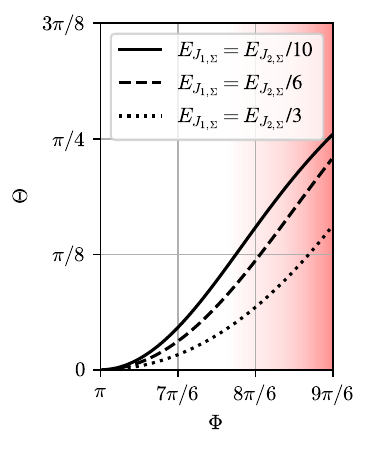}
            };
    
        \node at ($ (A) + (-2.45, 2.9) $) {\footnotesize \textbf{(a)}};
        \node at ($ (A) + (-1.5, -0.55) $) {\footnotesize \textbf{(b)}};
        \node at ($ (A) + (0.7, -0.55) $) {\footnotesize \textbf{(c)}};
        \node at ($ (B) + (-3.15, 2.4) $) {\footnotesize \textbf{(d)}};
        \node at ($ (B) + (-3.15, 0.45) $) {\footnotesize \textbf{(e)}};
        \node at ($ (B) + (-3.15, -1.5) $) {\footnotesize \textbf{(f)}};
        \node at ($ (C) + (-2.55, 2.9) $) {\footnotesize \textbf{(g)}};
    \end{tikzpicture}
    \caption{
        Flux-based control of a $\pi$-SQUID.
        Calculated numerically for $E_{J_{2, \Sigma}}/E_C = 30$, $E_{J_{1, \Sigma}}/E_{J_{2,\Sigma}} = 0.1$, $E_C/h = 200$ MHz, $n_g = 0$, and $d_1=d_2=0.05$.
        (a) Energy spectrum as a function of flux $\Phi$. 
        (b) Sketch of an adiabatic pulse schedule.
        (c) Sketch of a diabatic pulse schedule.
        (d) Larmor precession around the $z$-axis obtained by adiabatically splitting the qubit states.
        (e,f) Precession around a tilted axis obtained by diabatically pulsing away from half-flux.
        (g) Tipping angle $\Theta$ of the diabatic evolution as a function of flux $\Phi$.
        The red region indicates where the two-level approximation for a diabatic gate breaks down.
        Diabatic pulses further into this region increase the probability of leakage via Landau-Zener transitions.
    } 
    \label{fig:control}
\end{figure*}

To give some insight into this behaviour we obtain an approximate expression for the ground state splitting around half-flux. We can rewrite the $\pi$-SQUID Hamiltonian from~\cref{eq:circuit,eq:pi-SQUID} as $\hat H = \hat H_0 + \hat V $,
where the ideal $\pi$-periodic Hamiltonian is
\begin{equation}\label{eq:pi-periodic}
    \hat H_0 = 4E_C \left( \hat n - n_g \right)^2 - E_{J_2}(\Phi) \cos{2 \hat \varphi} ,
\end{equation}
and the $2\pi$-periodic term which perturbs it is
\begin{equation}\label{eq:2pi-periodic}
    \hat V = - E_{J_1}(\Phi) \cos{\left[\hat \varphi - \varphi_0(\Phi) \right]}.
\end{equation}
The Hamiltonian $\hat H$ will have two near-degenerate ground states, $\ket{\psi_0}$ and $\ket{\psi_1}$, localised in each of the two potential minima.
Since $E_{J_2}(\Phi)$ is negative at half-flux, these minima are located at $\varphi=\pm\pi/2$. 
For $E_{J_{1,\Sigma}}=0$ the eigenstates will be the symmetric and antisymmetric superpositions of the these localised states due to tunnelling. 
However even for very small $E_{J_{1,\Sigma}}$ the energies in each well become slightly different and the two lowest energy eigenstates are localised, as can be seen in figure~\cref{fig:spectrum}(b). 
This allows us to estimate the energy difference between the two lowest energy levels by making a harmonic approximation around each of the two wells and calculating energies up to first order in perturbation theory.
This is detailed in~\cref{sec:qubit-splitting}.
We find that the qubit energy splitting is
\begin{equation}
    E_q(\Phi) \approx 2 E_{J_1}(\Phi) \sin{\left[\varphi_0(\Phi)\right]} \exp{\left( -\sqrt{\frac{E_C}{8|E_{J_2}(\Phi)|}} \right)}.
\end{equation}
The extensive sweet spot is well described by this formula. 
The final, exponential factor is close to unity in the transmon-like regime around half-flux where $E_C \ll |E_{J_2}(\Phi)|$. 
The factor $E_{J_1}(\Phi) \sin{\left[\varphi_0(\Phi)\right]}$ proves to be nearly constant as a function of $\Phi$ in the region around $\Phi=\pi$, as can be seen in the qubit splitting plotted in figure~\cref{fig:spectrum}(d). 
This effect is enhanced for symmetric $\pi$-SQUIDs, with $d_1,d_2\approx0$, see~\cref{eq:phases,eq:relative-phase}.

Operating near half-flux suppresses the $2\pi$-periodicity to a surprising extent. 
In this region, a $\pi$-SQUID qubit has a small energy splitting and an extremely flat flux dispersion, providing strong protection against flux-based dephasing. 
By tuning to this regime, we can regain much of the protection afforded by an ideal $\pi$-periodic qubit, even with considerable single Cooper pair tunneling. 
In the following section, we show that universal control of a $\pi$-SQUID is possible via adiabatic and diabatic tuning of $\Phi$ around half-flux.



\section{Flux-based control}\label{sec:gates}

A $\pi$-SQUID qubit is well-suited to flux-based control due to its extended sweet spot.
Here, we describe an approach to qubit control that is analogous to the control of single-triplet spin qubits~\cite{petta2005coherent}.
Specifically, we describe set of adiabatic and diabatic pulses that provide a rotation around the qubit $z$-axis, and some other flux-dependent axis, respectively.

Adiabatic gates provide a straightforward means of $z$-axis rotation.
The basic principle is to increase the qubit frequency splitting $E_q$ between the logical states $\ket{\psi_0}$ and $\ket{\psi_1}$.
An example of an adiabatic pulse schedule is shown in~\cref{fig:control}(b).
Starting from $\Phi=\pi$, the flux is slowly ramped to some desired value of $\Phi$ where the splitting $E_q(\Phi)$ is larger, and, after waiting some time $\delta t$, the flux is slowly ramped back to $\Phi=\pi$.
This achieves a rotation around the $z$-axis $R_z(\theta)$, where $\theta = E_q \delta t / \hbar$. 
As an example, we have plotted the return probability $P_+$ for $\ket{\psi_+} = (\ket{\psi_0} + \ket{\psi_1}) / {\sqrt{2}}$, as a function of time in~\cref{fig:control}(d).
Faster evolution around $z$ requires pulsing further toward the unprotected region at $\Phi = 2\pi$.
We plot the qubit splitting $E_q(\Phi)$ for several values of $E_{J_{1,\Sigma}}$ in~\cref{fig:spectrum}(d).
However, pulsing further than $\Phi=3\pi/2$ is inadvisable, as it requires extremely slow pulses to navigate the small avoided crossing with higher-energy states, as seen in~\cref{fig:control}(a).

Depending on the total amplitude of single Cooper pair tunneling $E_{J_{1,\Sigma}}$, these adiabatic gates may not provide much more advantage than simply waiting for Larmor precession.
Nonetheless, they will be advantageous for more ideal $\pi$-junctions, which have smaller residual $2\pi$-tunneling. 

Rotation around any other axis of the Bloch sphere is sufficient to achieve universal single-qubit control.
We propose a diabatic gate schedule, an example of which is shown in~\cref{fig:control}(c).
Starting from $\Phi=\pi$, the flux is rapidly tuned to a desired value of $\Phi$.
The logical basis states $\ket{\psi_0}$ and $\ket{\psi_1}$ are no longer eigenstates of the system, and thus in the Bloch sphere picture they evolve around an axis that is not aligned with the $z$-axis.
After waiting some time, the flux is rapidly tuned back to $\Phi=\pi$.
This results in a rotation $R_\Theta(\theta)$ around some axis with tipping angle $\Theta$ from the $z$-axis.
We plot two examples of diabatic pulse evolution in~\cref{fig:control}(e,f).
These correspond to two different values of flux, as noted in~\cref{fig:control}(a).
Pulsing further from half-flux leads to both a larger tipping angle, and slightly quicker evolution.
We plot the tipping angle $\Theta$ as a function of $\Phi$ in~\cref{fig:control}(g).
Care should be taken not to pulse near to $\Phi=3\pi/2$, where the gap between the logical states and excited states is minimal.
Doing so runs the risk of incurring leakage errors due to Landau-Zener transitions.

By concatenating rotations around the $z$-axis and the $\Theta$-tilted axis, an arbitrary single-qubit rotation can be obtained.
These operations are possible due to the flux-tunable superposition of the $\pi$-periodic and $2\pi$-periodic potentials.
Therefore, flux-tunable control schemes will not work in the far-ideal limit where the $2\pi$-periodic contribution is negligible.
At this point, alternative schemes will be required. 

Aside from control, the flux-tunability of the $\pi$-SQUID potential can also be exploited for state preparation and measurement.
At the points $\Phi=\pi/2$ and $\Phi=3\pi/2$, the $\pi$-periodicity is highly suppressed, the $2\pi$-periodicity amplified and thus $E_{J_1}(\Phi) \gg E_{J_2}(\Phi)$.
For these values of flux, the circuit behaviour will most strongly resemble a conventional superconducting charge qubit.
For our choice of parameters, the spectrum in~\cref{fig:spectrum}(c) resembles a Cooper pair box at $\Phi=\pi/2$, with a single ground state, and pairs of excited states (note that we have set $n_g=0$).
A straightforward approach to preparation and measurement is to pulse to the regime of minimal protection, and utilise standard charge qubit techniques for initialisation and readout~\cite{Wallraff_2004}.
For preparation, the system will need to be adiabatically tuned from $\Phi=\pi/2$ to $\Phi=\pi$ after the initialisation.
Similarly, for measurement, the system will need to be adiabatically tuned from $\Phi=\pi$ and $\Phi=\pi/2$ and then readout.
This conversion between phase-like eigenstates and charge-like eigenstates is analogous to the spin-to-charge conversion used in singlet-triplet qubits for preparation and measurement.

This completes the single-qubit toolkit for non-ideal $\pi$-SQUIDs.
The flux-tunable bias of the double-well potential can be used for universal single-qubit gates, and supplemented with conventional superconducting charge qubit protocols to achieve state preparation and measurement.

\section{Holonomic Gates}\label{sec:holonomic}


In this section we consider the performance of the holonomic gate introduced in Ref.~\cite{Klots_2021} for finite but small values of $E_{J_{1,\Sigma}}$, since this  approach is an alternative to our preferred approach to performing universal operations.

\begin{figure}
    \centering
    \begin{tikzpicture}
        \coordinate (A) at (-2.1, 0);
        \coordinate (B) at (2.1, 0);
        \coordinate (C) at (0, -4.25);
    
        \node at (A) {
            \includegraphics[
                trim={0 0 15 0}, clip,
                scale=0.62
                ]{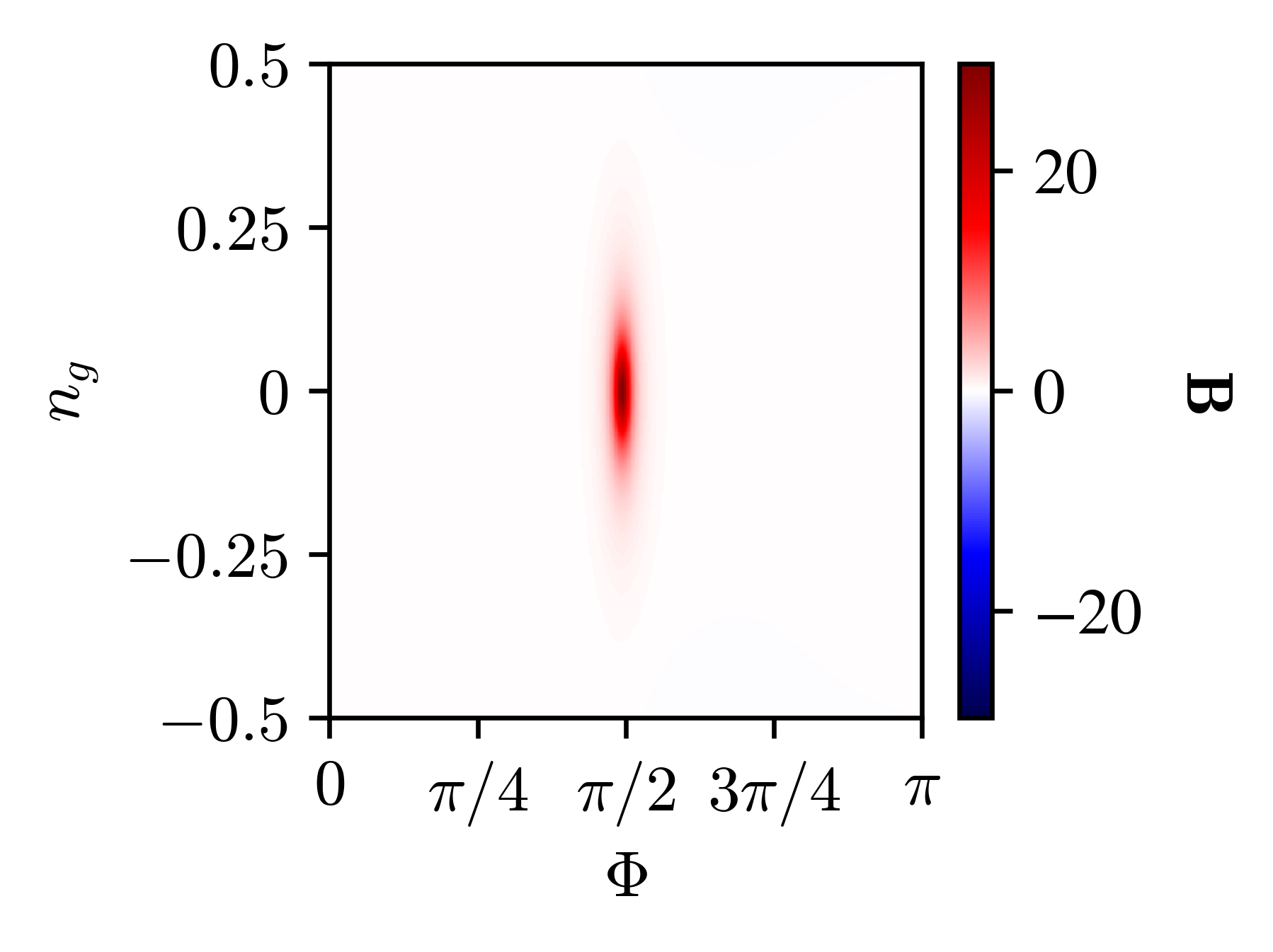}
            };
        \node at (B) {
            \includegraphics[
                trim={0 0 15 0}, clip,
                scale=0.62
                ]{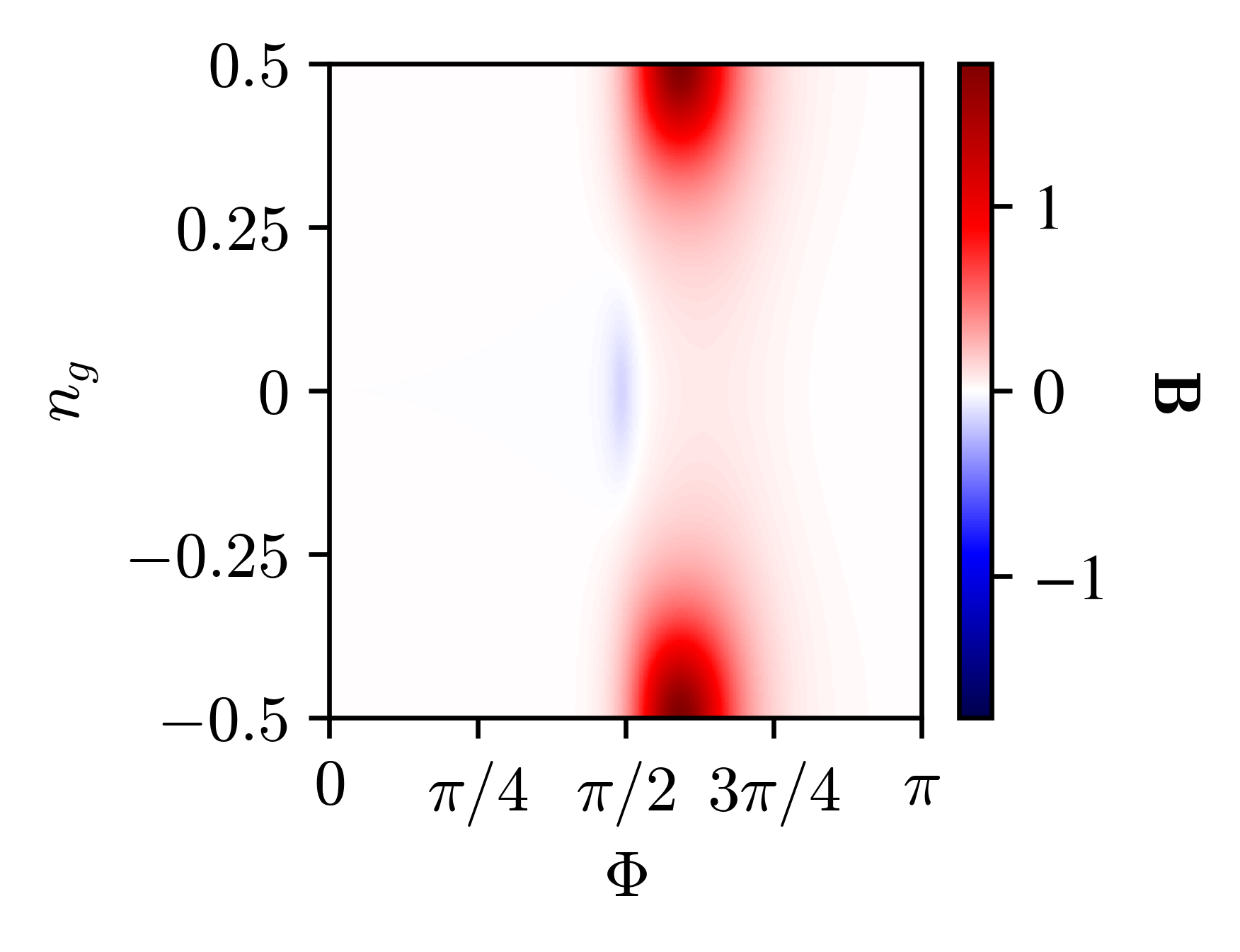}
            };
        \node at (C) {
            \includegraphics[scale=0.8]{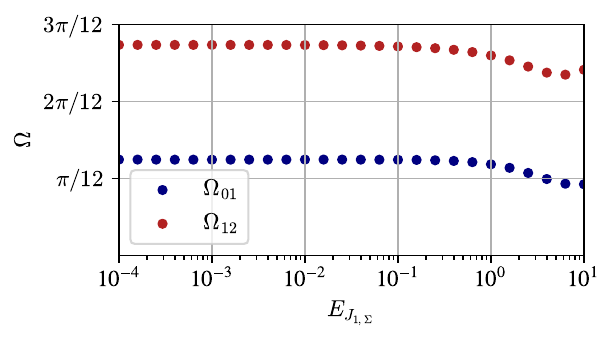}
            };

        \node at ($ (A) + (0.11, 1.8) $) {\footnotesize ${B}_{12}$};
        \node at ($ (B) + (0.11, 1.8) $) {\footnotesize ${B}_{01}$};
        
        \node at ($ (A) + (-1.9, 1.5) $) {\footnotesize \textbf{(a)}};
        \node at ($ (B) + (-1.9, 1.5) $) {\footnotesize \textbf{(b)}};
        \node at ($ (C) + (-2.1-1.9, 1.85) $) {\footnotesize \textbf{(c)}};
    \end{tikzpicture}
    \caption{
        Berry curvature and phase.
        Calculated numerically for $E_{J_{2, \Sigma}}/E_C = 30$, $E_{J_{1, \Sigma}}/E_{J_{2,\Sigma}} = 0.1$, $E_C/h = 200$ MHz, $n_g = 0$, and $d_1=d_2=0.05$.
        (a) Berry curvature between $\ket{\psi_1}$ and the excited state $\ket{\psi_2}$. 
        (b) Berry curvature between the logical states $\ket{\psi_0}$ and $\ket{\psi_1}$. 
        (c) Berry phases $\Omega_{12}$ and $\Omega_{01}$ obtained by a holonomic gate, due to the curvatures in (a) and (b), respectively.
    }
    \label{fig:holonomic}
\end{figure}
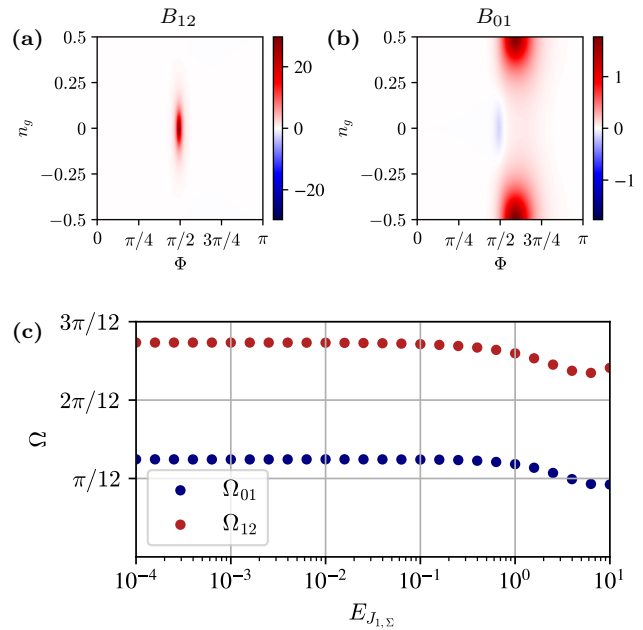

A holonomic gate utilises adiabatic traversal through a multi-dimensional parameter space to generate a geometric phase. 
This is a natural way to generate a relative phase in a qubit with degenerate logical states without lifting its protection.
As explained in~\cite{Klots_2021} a holonomic phase gate can be achieved in an ideal $\pi$-SQUID by varying the flux $\Phi$ and offset charge $n_g$, see~\cref{eq:circuit,eq:pi-SQUID}.

Any closed path in the space $\{\Phi, n_g\}$ generates a geometric phase for each eigenstate of the Hamiltonian that is dependent on the integral of the Berry curvature in the enclosed region. In this two dimensional parameter space the curvature is described by a single number.
The contribution to the curvature for a state $\ket{\psi_k}$ due to the state $\ket{\psi_l}$ is given by 
\begin{equation}
    B_{kl} = -\mathfrak{Im}
    \left[
\frac{\sum_{i,j=\Phi,n_g}\epsilon_{ij}\bra{\psi_k}\partial_i\hat{H}\ket{\psi_l}  \bra{\psi_l} \partial_{j} \hat{H}\ket{\psi_k} }{(E_k - E_l)^2} 
    \right].
\end{equation}
where $E_k$ and $E_l$ are the corresponding energy eigenvalues and $\epsilon_{ij}$ is the antisymmetric tensor. 
The Berry phase for a state $k$ due to moving around an closed path $\mathcal{C}$ in the parameter space is  
\begin{equation}
    \Omega_k=\sum_l\Omega_{kl} = \sum_{l}\oint_{\mathcal{C}}B_{kl}(\Phi,n_g)\: d\Phi\: dn_g.
\end{equation}

The proposal of~\cite{Klots_2021} is to perform a phase gate by varying flux and offset charge around a peak in curvature $B_{12}$ at $\Phi=0$ and $n_g=\frac12$, as shown in~\cref{fig:holonomic}(a).
By traversing the perimeter of the plotted region, where the $\pi$-SQUID ground states are degenerate, the state $\ket{\psi_1}$ accrues a phase due to an avoided crossing with the excited state $\ket{\psi_2}$.
This geometric phase results in a rotation around the $z$-axis since $\ket{\psi_0}$ is accrues no geometric phase.

However, in a $\pi$-SQUID with residual $2\pi$-periodicity, the degeneracy between ground states is broken and the qubit states are coupled by the $E_{J_1}(\Phi)$ term in the Hamiltonian.  
This produces a curvature between states $\ket{\psi_0}$ and $\ket{\psi_1}$, as shown in~\cref{fig:holonomic}(b).
This curvature is localized near $n_g = \pm \frac12$, which is directly in the path of the protected gate. 
Navigating this residual curvature is a concern for implementing holonomic gates in non-ideal $\pi$-SQUIDs.
Also, due to this additional coupling the path chosen in \cite{Klots_2021} passes through an avoided crossing of the qubit states at $n_g=\pm$ and close to $\Phi=\pi/2$. 
To remain adiabatic the path will have to be deformed to avoid this location. 
However, due to the nonzero curvature in this region the gate will be much more sensitive to the precise path in $\Phi$ and $n_g$.

In~\cref{fig:holonomic}(c) we show the phase accrued from a holonomic gate as described above. 
We plot both the phase $\Omega_{12}$ accrued between $\ket{\psi_1}$ and $\ket{\psi_2}$, due to the curvature shown in~\cref{fig:holonomic}(a), and the phase $\Omega_{01}$ accrued between $\ket{\psi_0}$ and $\ket{\psi_1}$, due to the curvature shown in~\cref{fig:holonomic}(b).
For moderate values of $E_{J_{1,\Sigma}}$, the overall phase will be subject to flux noise, which can enter through $E_{J_1}(\Phi)$.
For lower values of $E_{J_{1,\Sigma}}$, the phase is constant, but the ideal phase will be $\Omega_{12}$ reduced due to the phase $\Omega_{01}$.

For these reasons we prefer our approach of pulsed gates based on singlet-triplet qubit operations to the use of holonomic gates when there is nonzero $E_1(\Phi)$. 
One option to mitigate these effects might be to design pulses that pass through the qubit transition at $n_g=\pm \frac12$ diabatically while remaining adiabatic with respect to the transition between $\ket{\psi_1}$ and $\ket{\psi_2}$ used in the original proposal.



\section{Conclusion}\label{sec:conclusions}

In summary, we have demonstrated the principles for flux-based control of a non-ideal $\pi$-SQUID qubit. 
Our scheme is broadly applicable to many protected superconducting circuits as well as qubits based hybrid superconductor-semiconductor nanowire junctions.
The $\pi$-SQUID allows for flux-tunable qubit state protection, and features a remarkably broad region around half-flux where the effect of parasitic $2\pi$-periodicity of the constituent circuits can be minimised.
These findings are supported by both numerical and analytical calculations.
We have described the basic principles of flux-based control, state preparation, and measurement, all while operating in the regime of enhanced protection.
We believe that experimentally-approachable methods of control for protected qubits are important in the near-term regime where circuits are too well-protected to control using gates based on oscillatory fields, but not protected enough to see improved coherence and error rates.

\section*{Acknowledgements}

We acknowledge support from Australian Research Council via the Centre of Excellence in Engineered Quantum Systems (CE170100009) and the US Army Research Office (W911NF-23-10092).
We acknowledge the traditional owners of the land on which this work was undertaken at the University of Sydney, the Gadigal people of the Eora Nation.
TBS would like to thank Xanthe Croot for insightful discussions.

\appendix

\section{Formulae for a $\pi$-SQUID}\label{app:pi-SQUID-derivation}

The Josephson Hamiltonian
\begin{equation}
    \begin{aligned}
        \hat{H}_J &= - \sum_{i = a,b} \left( E_{J_{1,i}} \cos{\hat\phi_i} + E_{J_{2,i}} \cos{2\hat\phi_i} \right),
    \end{aligned}
\end{equation}
may be rewritten in terms of the total phase difference $\hat\varphi = (\hat\phi_a + \hat\phi_b)/2$:
\begin{equation}
    \begin{aligned}
        \hat H_J ={}& - E_{J_1} (\Phi) \cos{\left[\hat\varphi - \varphi_1(\Phi) \right]}
        \\ {}& - E_{J_2} (\Phi) \cos{\left[ 2 \hat\varphi - \varphi_2 (\Phi)  \right]},
    \end{aligned}
\end{equation}
where the effective Josephson energies are
\begin{align}\label{eq:josephson-energies}
    E_{J_1}(\Phi) &= E_{J_{1,\Sigma}} \cos{(\Phi/2)} \sqrt{1 + d_1^2 \tan^2{(\Phi/2)}} ,
    \\ E_{J_2}(\Phi) &= E_{J_{2,\Sigma}} \cos{(\Phi)} \sqrt{1 + d_2^2 \tan^2{(\Phi)}} ,
\end{align}
the phase shifts are determined by
\begin{align}\label{eq:phases}
    \varphi_1 (\Phi) &= \arctan{\left[ d_1 \tan{(\Phi/2)} \right]} ,
    \\ \varphi_2 (\Phi) &= \arctan{\left[ d_2 \tan{(\Phi)} \right]} ,
\end{align}
the total tunneling amplitudes are denoted
\begin{align}
    E_{J_{1,\Sigma}} &= E_{J_{1,a}} + E_{J_{1,b}},
    \\ E_{J_{2,\Sigma}} &= E_{J_{2,a}} + E_{J_{2,b}},
\end{align}
and the relative asymmetry in the junctions are 
\begin{align}
    d_1 &= \frac{E_{J_{1,a}} - E_{J_{1,b}}}{E_{J_{1,a}} + E_{J_{1,b}}},
    \\ d_2 &= \frac{E_{J_{2,a}} - E_{J_{2,b}}}{E_{J_{2,a}} + E_{J_{2,b}}}.
\end{align}
The phase shifts $\varphi_1(\Phi)$ and $\varphi_2(\Phi)$ cannot be removed via a change of variables.
For convenience, we define the relative phase shift  
\begin{equation}\label{eq:relative-phase}
    \varphi_0(\Phi) = \varphi_1(\Phi) - \frac{\varphi_2(\Phi)}{2},
\end{equation}
and rewrite the Josephson Hamiltonian as
\begin{equation}
    \hat H_J = - E_{J_1}\left( \Phi \right) \cos{\left[\hat \varphi - \varphi_0\left( \Phi \right) \right]} - E_{J_2}\left( \Phi \right) \cos{2 \hat \varphi} .
\end{equation}
by shifting the phase operator $\hat\varphi \rightarrow \hat\varphi + \varphi_2(\Phi)/2$.

\section{Approximation for the qubit energy splitting}\label{sec:qubit-splitting}

We can rewrite the $\pi$-SQUID Hamiltonian as $\hat H = \hat H_0 + \hat V $,
where the ideal $\pi$-periodic Hamiltonian is
\begin{equation}
    \hat H_0 = 4E_C \left( \hat n - n_g \right)^2 - E_{J_2}(\Phi) \cos{2 \hat \varphi} ,
\end{equation}
and the $2\pi$-periodic term which perturbs it is
\begin{equation}
    \hat V = - E_{J_1}(\Phi) \cos{\left[\hat \varphi - \varphi_0(\Phi) \right]}.
\end{equation}
The Hamiltonian has two degenerate ground states, $\ket{\psi_0}$ and $\ket{\psi_1}$, localised at each of the two potential minima. 
Since $E_{J_2}(\Phi)$ is negative at half-flux, these minima are located at $\varphi=\pm\pi/2$. We define a new phase operator $ \delta \hat\varphi = \hat \varphi - \varphi_* $ that is shifted by an amount $\varphi_*$. If we shift to the well at $\varphi_*=\pi/2$, we obtain
\begin{align}
    \hat H_0 &= 4E_C \left( \hat n - n_g \right)^2 + E_{J_2}(\Phi) \cos{2 \delta \hat \varphi} ,
    \\ \hat V &= E_{J_1}(\Phi) \sin{\left[ \delta \hat\varphi - \varphi_0(\Phi) \right]} .
\end{align}
Since $E_C \ll |E_{J_2}(\Phi)|$, the Hamiltonian $\hat H_0$ is approximately harmonic around $\delta\varphi=0$.
Thus, the ground state $\ket{\psi_0}$ is localised at $\varphi_*=\pi/2$ and satisfies 
\begin{align}
    \langle \delta \hat{\varphi} \rangle &= 0  ,
    \\ \langle \delta {\hat\varphi}^2 \rangle &= \sqrt{ E_{C} / 2|E_{J_2}(\Phi)|} ,
\end{align}
which can be used to show that
\begin{align}
    \langle \sin \delta \hat{\varphi} \rangle &= 0  ,
    \\ \langle \cos{\delta \hat{\varphi}} \rangle &= \exp{( -\sqrt{E_C/8|E_{J_2}(\Phi)|} )} .
\end{align}
The first-order energy shift to $\ket{\psi_0}$ due to $\hat V$ is
\begin{equation}
    \langle \hat V \rangle_0 \approx E_{J_1}(\Phi) \sin{\left[\varphi_0(\Phi)\right]} \cos{\delta \hat{\varphi}}
\end{equation}
Using an identical argument for the other well located at $\varphi_*=-\pi/2$, we find that the first-order energy shift to the other ground state $\ket{\psi_1}$ is $\langle \hat V \rangle_1 = -\langle \hat V \rangle_0$.
Therefore, the qubit energy splitting around half-flux is
\begin{equation}
    E_q(\Phi) \approx 2 E_{J_1}(\Phi) \sin{\left[\varphi_0(\Phi)\right]} \exp{\left( -\sqrt{\frac{E_C}{8|E_{J_2}(\Phi)|}} \right)}.
\end{equation}

\bibliography{bibliography}

\end{document}

%% file: figures/qubit.tex
\input{figures/circuits}

\coordinate (A) at (0, 0);
\begin{scope}[shift=(A), rotate=0, scale=0.75]

    \newcommand\QubitHeight{2.5}
    \newcommand\QubitWidth{2}
    
    \draw [line width=\WireLinewidth, color=WireColor, fill=Background, rounded corners=\WireCorners]
        (-\QubitWidth/2, -\QubitHeight/2) rectangle (\QubitWidth/2, \QubitHeight/2);


    \coordinate (C) at (\QubitWidth/2, 0);
    \Capacitor{(C)}{90};

    \node at ($ (C) + (1, 0) $) {$C_{\phantom{g}}$};


    \coordinate (S) at (-\QubitWidth/2, 0);
    \newcommand\SQUIDHeight{1.5}
    \newcommand\SQUIDWidth{1.25}

    \begin{scope}[shift=(S)]
        \draw [line width=\WireLinewidth, color=WireColor, fill=Background, rounded corners=\WireCorners]
            (-\SQUIDWidth/2, -\SQUIDHeight/2) rectangle (\SQUIDWidth/2, \SQUIDHeight/2);

        \coordinate (P1) at (-\SQUIDWidth/2, 0);
        \coordinate (P2) at (\SQUIDWidth/2, 0);
    
        \PJunction{(P1)}{0};
        \PJunction{(P2)}{0};

        \node at ($ (P1) + (-0.3, 0.5) $) {$a$};
        \node at ($ (P2) + (0.3, 0.5) $) {$b$};

        \coordinate (F) at (0, 0.2);
        \draw (F) [line width=\WireLinewidth, fill=WireColor] circle (0.05);

        \node at ($ (F) + (0, -0.4) $) {$\Phi$};
    \end{scope}


    \coordinate (G) at (0, -\QubitHeight/2-0.33);

    \draw (0, -\QubitHeight/2) [line width=\WireLinewidth] to (G);
    \draw (0, -\QubitHeight/2) [line width=\WireLinewidth, fill=WireColor] circle (0.08);
    
    \Ground{(G)}{0};


    \coordinate (N) at (0, \QubitHeight/2);

    \draw (N) [line width=\WireLinewidth, fill=WireColor] circle (0.08);


    \coordinate (VG) at (-\QubitWidth/4-\SQUIDWidth/4, \QubitHeight/2 + 1.2);
    \coordinate (CG) at (0, \QubitHeight/2 + 0.5);
    \coordinate (GG) at (-\QubitWidth/2-\SQUIDWidth/2, \QubitHeight/2 + 0.5);

    \draw [line width=\WireLinewidth, color=WireColor, rounded corners=\WireCorners]
        (GG)
        to ++(0, 0.7)
        to (VG) 
        to ++(\QubitWidth/4+\SQUIDWidth/4, 0) 
        to (0, \QubitHeight/2);

    \renewcommand\CapLength{0.5}https://www.overleaf.com/project/65160767f53724f604d391fd

    \Capacitor{(CG)}{90}
    \node at ($ (CG) + (0.75, 0) $) {$C_g$};
    
    \draw (VG) [line width=\WireLinewidth, fill=Background] circle (0.375);
    \node at (VG) {$V_g$};

    \Ground{(GG)}{0}
    




\coordinate (B) at (0, -3.5);

\begin{scope}[shift=(B)]

    \newcommand\SegLength{2}

    \coordinate (L) at (-\QubitWidth/2-\SQUIDWidth/2, 0);

    \begin{scope}[shift=(L)]
        \draw (0, \SegLength/2) [line width=\WireLinewidth] to (0, -\SegLength/2);
        \draw (0, \SegLength/2) [line width=\WireLinewidth, fill=Background] circle (0.08);
        \draw (0, -\SegLength/2) [line width=\WireLinewidth, fill=Background] circle (0.08);

        \PJunction{(0, 0)}{0};
    \end{scope}

    \node at ($ (L) + (0.9, 0) $) {\Large $=$};


    \newcommand\LoopHeight{1.2}
    \renewcommand\LoopWidth{1}

    \coordinate (R) at (\QubitWidth/2, 0);

    \begin{scope}[shift=(R)]
        \draw (0, \SegLength/2) [line width=\WireLinewidth] to (0, -\SegLength/2);
        \draw (0, \SegLength/2) [line width=\WireLinewidth, fill=Background] circle (0.08);
        \draw (0, -\SegLength/2) [line width=\WireLinewidth, fill=Background] circle (0.08);
        
        \draw [line width=\WireLinewidth, color=WireColor, fill=Background, rounded corners=\WireCorners]
            (-\LoopWidth/2, -\LoopHeight/2) rectangle (\LoopWidth/2, \LoopHeight/2);

        \coordinate (PJ) at (\LoopWidth/2, 0);
        \coordinate (JJ) at (-\LoopWidth/2, 0);

        \PJunctionIdeal{(PJ)}{0};
        \JJunctionIdeal{(JJ)}{0};

        \draw [<->, distance=15, out=-45, in=45] 
            ($ (PJ) + (0.33, \LoopHeight/2) $) 
            to ++(0, -\LoopHeight);

        \draw [<->, distance=15, out=-135, in=135] 
            ($ (JJ) + (-0.33, \LoopHeight/2) $) 
            to ++(0, -\LoopHeight);

        \node at ($ (PJ) + (0.8, -\LoopHeight/2) $) {$4e$};
        \node at ($ (JJ) + (-0.8, -\LoopHeight/2) $) {$2e$};

    \end{scope}
\end{scope}

\end{scope}`

\node at (4.3, -0.75) {\includegraphics[width=0.6\linewidth]{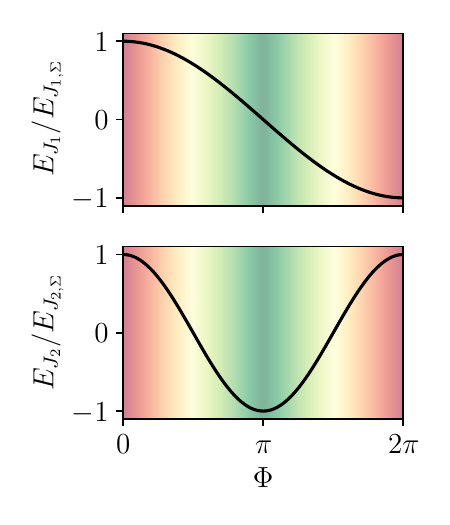}};

\node at (-1.75, 1.9) {\footnotesize \textbf{(a)}};
\node at (-1.75, -2) {\footnotesize \textbf{(b)}};
\node at (2.225, 1.9) {\footnotesize \textbf{(c)}};
\node at (2.225, -0.55) {\footnotesize \textbf{(d)}};